\documentclass[journal]{IEEEtran}
\usepackage[explicit]{titlesec}
\usepackage{graphicx}
\usepackage{color, soul}
\usepackage{epstopdf}
\usepackage{subfigure} 
\usepackage{algorithmic}
\usepackage{amsmath}
\usepackage{amsthm}
\usepackage[left]{lineno}
\usepackage{cite}

\usepackage{bbm}
\usepackage{flushend}
\usepackage[colorlinks, linkcolor=blue,citecolor=blue]{hyperref}
\usepackage{amssymb}

\newcommand{\RNum}[1]{\uppercase\expandafter{\romannumeral #1\relax}}

\usepackage{stfloats}

\usepackage{color}
\usepackage{tikz,xcolor}

\definecolor{lime}{HTML}{A6CE39}

\titlespacing{\section}{0pt}{1.2ex plus .0ex minus .0ex}{.3ex plus .0ex}
\titlespacing{\subsection}{0pt}{1.2ex plus .0ex minus .0ex}{.3ex plus .0ex}

\makeatletter
\DeclareRobustCommand{\orcidicon}{%
	\begin{tikzpicture}
	\draw[lime, fill=lime] (0,0) 
	circle [radius=0.16] 
	node[white] {{\fontfamily{qag}\selectfont \tiny ID}};    \draw[white, fill=white] (-0.0625,0.095)
	circle [radius=0.007];    \end{tikzpicture}
	\hspace{-2mm}}
\foreach \x in {A, ..., Z}{%
\expandafter\xdef\csname orcid\x\endcsname{\noexpand\href{https://orcid.org/\csname orcidauthor\x\endcsname}{\noexpand\orcidicon}}
}

\newcommand*\bigcdot{\mathpalette\bigcdot@{.5}}
\newcommand*\bigcdot@[2]{\mathbin{\vcenter{\hbox{\scalebox{#2}{$\m@th#1\bullet$}}}}}
\makeatother

\usepackage[linesnumbered,ruled,vlined]{algorithm2e}

\definecolor{mydarkblue}{RGB}{65, 105, 225}

\begin{document}
\title{Towards {Goal-Oriented Semantic} Communications: New Metrics, Framework, and Open Challenges}
\author{Aimin Li\orcidA{}, 
	Shaohua Wu\orcidB{}, 
	\emph{Member, IEEE,}
	Siqi Meng\orcidC{},
	Rongxing Lu\orcidH{}, \emph{Fellow IEEE},\\
	Sumei Sun\orcidG{}, \emph{Fellow IEEE},
	and Qinyu Zhang\orcidF{}, 
	\emph{Senior Member, IEEE}
}

\maketitle

\begin{abstract}
Since Shannon's pioneering masterpiece which established the foundation of modern information theory, the design target of communications has long been promising bit-level message reconstruction and achieving \emph{shannon capacity}, which neglects the semantics and effectiveness aspects of information. Nevertheless, the recent development of wireless technologies and the spurt of deep learning (DL) techniques allow us to reclaim the meaning/usefulness aspect in the design of future 6G communication paradigms, where {goal-oriented communication} is becoming a trend. \emph{Age of Information} (AoI), a well-known metric that captures the \emph{significance} of information by recording the time elapsed from the generation time slot, has been extended to various variants, such as \emph{Value of Information} (VoI), \emph{Urgency of Information} (UoI), \emph{Age of Incorrect Information} (AoII), and etc. While each of them proposes novel ways to measure the semantics/effectiveness aspect of information, there is not yet a unified framework encompassing all of them. To this end, we propose a novel tensor-based approach, the \emph{Goal-oriented Tensor} (GoT), to unify them, which also allows more flexible and fine-grained goal characterizations. {Following the proposed GoT, we architect a holistic goal-oriented semantic communications framework, in which semantics perception, dissemination, and control-plane decision making collaboratively work towards realizing specific goals.} Finally, we outline several open challenges to fulfill the vision of the GoT framework.
\end{abstract}


\IEEEpeerreviewmaketitle

\section{Introduction} 
The modern prosperity of communication technologies is primarily standing on the cornerstone built by Shannon's Classical Information Theory in 1948 \cite{6773024}. Since its inception, a primary goal in the field of communication has been to approach Shannon's theoretical capacity for transmission. Today, the advanced 5G technology has nearly achieved this goal. Technically, advances like massive multiple-input multiple-output (MIMO), and millimeter wave communications have enabled a plethora of services with ultra reliability, low latency, and high data rate. At the physical layer, the combination of Low-Density Parity-Check (LDPC) codes and Polar codes also represents a breakthrough in realizing low-complexity near-Shannon-limit coding-decoding pairs. Such technologies aim at building perfect bit-by-bit information pipelines between transceivers, which follow Shannon's legacy precisely.

Looking forward, we could anticipate a massive data boom in the forthcoming data-driven world. On the one hand, non-terrestrial nodes are being deployed throughout the space to assure seamless coverage and ubiquitous connectivity. This expansion amplifies bit generation sources. Meanwhile, data types are becoming diversified and versatile, encompassing Virtual Reality (VR), high-resolution video streams, and multi-modal sensing data. As the raw data demands massive amounts of bits to characterize, the limited communication resources are further burdened. Inevitably, the traditional paradigm will suffer from a bottleneck due to limited communication resources that are not aligned with the requirements for bit-oriented ubiquitous data reconstruction. 

\subsection{Beyond Bit-Oriented Transmission}
To overcome the above challenges, Shannon's classical theory is in need of a paradigm shift, which can be traced back to Shannon's seminal work \cite{weaver}, where Shannon and Weaver categorized communications into three basic levels: \textit{technical} level, \textit{semantic} level, and \textit{effectiveness} level. \textit{Technical} level aims to accurately transmit data bits. \textit{Semantic} level aims to precisely convey the desired meaning. \textit{Effectiveness} level aims to transmit \emph{relevant} symbols such that the receiver makes effective decisions to affect conducts in the desired way. Both the \textit{semantic} level and the  \textit{effectiveness} level move forward to go beyond bit-oriented transmission, which have attracted extensive research interests recently.

Goal-oriented semantic communication represents an orchestration of the \textit{semantic} level and the \textit{effectiveness} level, and its key idea is prioritizing the sharing of goal-relevant semantic fragments, such that the decision making is accurate and the ultimate goal of the system is effectively achieved. 


\subsection{Overview of Semantic Communications}
There are two primary avenues that lead the development of semantic communications. The first avenue applies DL-learning techniques to extract semantic features and filter semantics-irrelevant redundancy within a specific packet, thus enhancing communication efficiency. The second avenue interprets semantics as its etymological meaning, the \textit{significance} of information (also known as \textit{priority} of information or \textit{relevance} of information).  

 
  {The first avenue} focuses on the engineering fulfillment of semantic compression and extraction, in which DL techniques play indispensable roles \textcolor{black}{to accurately extract semantic information within a specific packet}. Such a promising avenue arises from the interest in applying DL at the physical layer \cite{QinDLPLC},  \textcolor{black}{which facilitates a shift from bit reconstruction-oriented to semantic similarity-oriented transmission.} An earlier End-to-End (E2E) prototype for DL-based semantic communications is the Deep Joint Channel Source Coding (Deep JSCC) \cite{JSCC}, \textcolor{black}{which is used to extract (at the transmitter) and reconstruct (at the receiver) the semantics of an image.} Today, such a paradigm underpins a wide range of applications, such as text, video, and speech-based semantic communication. A more comprehensive view of such a track could be found in \cite[Sec. IV]{gunduz2022beyond} and \cite[Sec. III]{survey}.

{The second avenue} interprets semantics as ``\textit{significance}'', ``\textit{priority}'', or ``\textit{relevance}''\cite{uysal2022semantic}, which is also our focus in this article. From this perspective, intelligent networked systems can identify high-\textit{priority} messages and adaptively allocate more resources. \textcolor{black}{Notably, the determination of \emph{priority} deviates from traditional, probability-centric entropy measures. Instead, it leans on the ``\emph{relevance}'' attributed to the goal-oriented usefulness of message.} As noted in \cite{Nikolaos}, \emph{``Imagine two equally rare events, occurring with very low probability, one of which carries a major safety risk while the other is just a peculiarity. Although they provide the same high amount of information, the information conveyed by the first event is evidently of higher significance.''}, it holds great potential to design a semantics/goal-aware \emph{relevant filter} to slim down the information pipeline, where only packets that carry \emph{relevant} meaning for accomplishing the goal are prioritized for transmission. {Since the \emph{relevance} relates to the goal, the \emph{relevance} interpretation also bridges this avenue to the effectiveness level, motivating us to expand upon goal-oriented semantic communications from this viewpoint.}
	
 In the second avenue, the core challenge is to design metrics that could characterize the \emph{relevance} property. Age of Information (AoI), defined as the time elapsed since the generation of the most recent information update at the destination, is one typical metric that captures the \emph{relevance} property. In the AoI-oriented system, fresher messages are deemed more \emph{relevant} to the end-user, and thus prioritized for transmission. However, AoI does not perceive the content of the packet, nor the task at the receiver. To address this issue, improved non-linear variants of AoI, such as VoI \cite{kosta2017age}, AoII \cite{AoII}, UoI \cite{zheng2020urgency}, Cost of Actuation Error \cite{9551200}, and etc., have been developed. However, there is not yet a unified framework that encompass all these metrics. Moreover, the existing metrics do not directly characterize the desired goal. These issues motivate us to propose a unified, flexible, and directly goal-oriented metric that could assist in accurately capturing the underlying goal, facilitating decision-making, and ultimately improving effectiveness.

\subsection{Towards Goal-Oriented Semantic Communications} \label{IC}
Following the second avenue, this article addresses the challenge of metrics design. First, this article comprehensively reviews the existing metrics that capture the \emph{relevance} property in Section \ref{section II}. In this way, we reveal how these metrics manifest the effectiveness level. Then, upon examining the inherent relationships among these metrics, we are inspired to unify them in a more cohesive framework. This leads to the proposed Goal-oriented Tensor (GoT). We visualize examples to demonstrate how the GoT metric reduces to existing metrics that captures \emph{relevance} and how it enables more fine-grained and flexible goal characterizations. Consequently, abstract goals could be defined and quantified via GoT.

 Furthermore, this article illustrates that information traverses a perception-actuation \emph{life cycle} to achieve a certain goal. {Within this \emph{life cycle}, information initiates with its generation, goes through semantics extraction (the meaning-aware approach will enhance the effectiveness), coding \& modulation, data dissemination, and ends with its transformations into effective decisions (or control) towards goals. In this way, the \emph{relevance} of information is not solely tied to the packet's inherent meaning, but the ultimate usage of the meaning for achieving particular goals. 
 	
 Incorporating the \emph{life cycle}, this article then envisions a holistic framework of \textcolor{black}{goal-oriented semantic} communications, whereby \textcolor{black}{semantics} perception, dissemination, and control-plane decisions \textcolor{black}{are orchestrated in harmony with a shared goal characterized by GoT.} We consider an easy-to-follow case study, a real-time wireless fire monitor and rescue system, to illustrate the universality of our proposed framework. Through this example, we illustrate how the GoT describes the goal in real scenarios. The idea of \textcolor{black}{ goal and semantics-aware filter} (an open challenge proposed in \cite{uysal2022semantic} and \cite{Nikolaos}) is also initially achieved in this article.} 

\section{Capturing the Relevance of Information: Critical Metrics}\label{section II}
Recently, many metrics have been investigated to characterize the \emph{relevance} property of information. This section provides an overview of these metrics and discusses their influence on effectiveness.
\begin{table*}[htbp]
	\caption{\textcolor{black}{Metrics for Capturing the Importance of Information: A Review}}
	\centering
	\includegraphics[angle=0,width=1\textwidth]{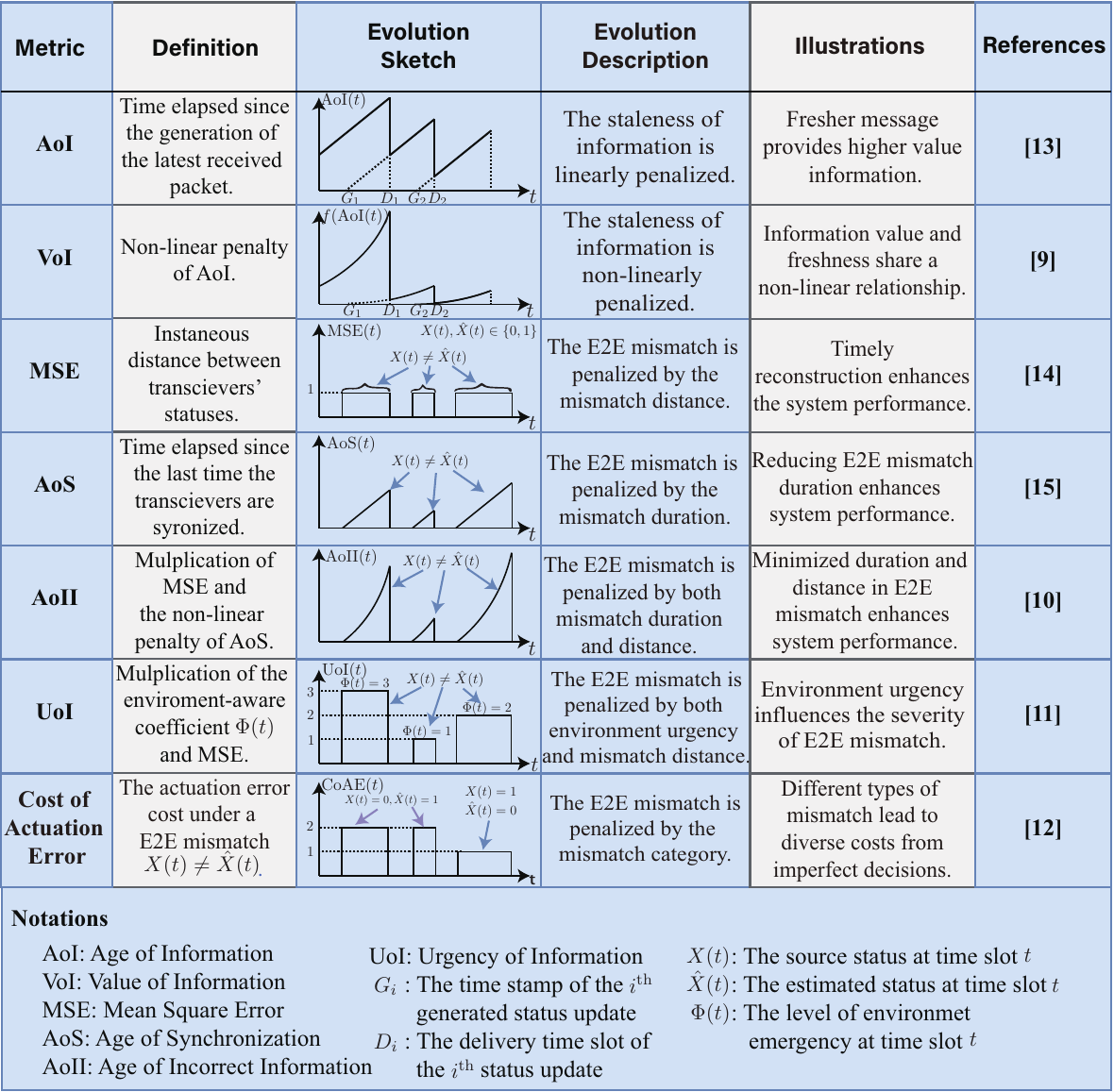}
	\label{review}
\end{table*}
\subsection{Age of Information (AoI)}

Age of Information (AoI), proposed in 2012 \cite{packetmagnament2}, is the first concrete and quantitative metric to characterize the \emph{relevance} of information. Intuitively known as \emph{freshness} of information, AoI was designed to fulfill the \emph{freshness} requirements for the proliferated machine-type communications, such as intelligent vehicle networks, industrial internet of things, and etc. Central to optimizations on AoI is a subtle yet vital consensus:  \emph{fresher messages contain more valuable information}. \textcolor{black}{This consensus is rooted in the understanding that \emph{fresh} information ensures more informed and goal-oriented decision-making.}

\subsection{Value of Information (VoI)}
VoI introduced a non-linear penalty of AoI to capture the degree of ``\emph{dissatisfaction}'' resulted by \emph{staleness} of information \cite{kosta2017age}. A preliminary approach to characterize this non-linear penalty is to classify the applications into basic levels in terms of their sensitivity to \emph{freshness} \cite{kosta2017age}. \emph{Freshness}-critical applications are penalized exponentially (see Table \ref{review} for an example), \emph{freshness}-insensitive ones are penalized logarithmically, and those neutral ones fall into linear penalties. \textcolor{black}{VoI bridges AoI with {effectiveness} by applying a non-linear utility function.}

\subsection{Mean Square Error (MSE)}
MSE is a well-known metric that characterizes the accuracy aspect of information. Defined as {\textit{Euclidean} distance between transceivers' statuses}, MSE is used for timely reconstruction-oriented communications. In the MSE-oriented design, a status at the source is deemed \emph{significant} if it considerably deviates from the estimated status at the monitor. When the sample policy is content-agnostic, the MSE can be expressed as a nonlinear function of AoI \cite{msee}, making it a specialized form of VoI. \textcolor{black}{MSE emphasizes the impact of the severity of E2E status mismatch on the effectiveness.}

\subsection{Age of Synchronization (AoS)}
As its name suggests, AoS measures how long has passed since the last time the transceivers synchronized their statuses \cite{zhongnewmetric}, addressing AoI's content-agnostic nature. Consider scenarios where the source undergoes frequent changes: even a ``\emph{fresh}'' packet with a low AoI might become obsolete if the true status of the source has significantly evolved. Conversely, a ``\emph{stale}'' packet can still offer accurate estimations if the source changes are minimal. \textcolor{black}{As shown in Table \ref{review}, AoS posits that the duration of E2E mismatch adversely impacts {effectiveness}. Thus, minimizing AoS improves the effectiveness.} 

\subsection{Age of Incorrect Information (AoII)}
\textcolor{black}{AoII is a new metric that orchestrates the strengths of MSE and AoS to measure \emph{relevance} property \cite{AoII}}. As shown in Table \ref{review}, AoII introduces three key innovations: First, AoII integrates AoS with MSE by multiplying the variants of them together, which incorporates both the duration and severity of E2E mismatch. Second, AoII incorporates a nonlinear penalty function predicated on AoS, drawing parallels to the penalty function in the Value of Information (VoI) relative to AoI. Third, AoII introduces a generalized error gap function, broadening the traditional Euclidean-distance-based MSE depictions. In this regard, \textcolor{black}{AoII addresses that both severity and duration of E2E statues mismatch affect the {effectiveness}.}

\subsection{\textcolor{black}{Cost of Actuation Error}}
\textcolor{black}{Previous metrics typically measure the system {effectiveness} indirectly. To address this issue, \emph{Cost of Actuation Error} emerges as the first metric to directly quantify the {effectiveness} at the point of actuation \cite{9551200}. This metric highlights the role of the actuator to affect the goal-achieving {effectiveness}: an E2E status mismatch will trigger an actuation error, hence inducing associated cost. As Table \ref{review} shows, this metric highlights the direct impacts of actuation error on {effectiveness}.}

\subsection{Urgency of Information (UoI)}
While previous metrics focus solely on the \emph{relevance} of the source, practical scenarios demonstrate that the surrounding environment also influences information relevance. For example, a self-driving car in complex situations like traffic jams would require more frequent status updates for safety, unlike when it's on a clear highway. Addressing this, UoI is the first metric to link environment factors with information relevance \cite{zheng2020urgency}. Particularly, UoI introduces an environment-aware weighted coefficient based on the environment urgency level. The final UoI formula is a product of this coefficient and the error gap function, as detailed in Table \ref{review}.

\section{One More Step Forward: the Goal-oriented Tensor}\label{sectionIII}
The above metrics each examine effectiveness from varied indirect perspectives. In this section, we introduce a unified metric designed to gauge the \emph{relevance} of information in a goal-oriented manner and seek to bring coherence to these various metrics.

\subsection{Existing Metrics: Inherent Connection}\label{discovery}
From Section \ref{section II} and Table \ref{review}, the existing metrics share two fundamental components: $i$) a \textbf{content-aware} error cost function $g(X(t),\hat{X}_t)$ and $ii$) a \textbf{content-independent} weighted coefficient $\Phi(t)$. The former one describes the real-time E2E distance (like MSE or other error gap function) or mismatch cost (like \emph{Cost of Actuation Error}). The latter one exerts a multiplicative effect on the error cost function. In this way, the inherent connection among the existing metrics is that they are all contingent upon a triple-tuple consisting of the status $X\left(t\right)$, the estimated status $\hat{X}\left(t\right)$, and the context $\Phi\left(t\right)$. This connection motivates us to propose a tensor-based approach to unify the existing metrics. 
\subsection{Goal-Oriented Tensor: A Unified Metric}

\textcolor{black}{Given that all existing metrics are based on a triple-tuple, a unified approach is natural: 
\emph{Unify the existing metrics as a 3-dimensional tensor, with each dimension corresponding to a tuple element.}}
\noindent In this subsection, We first show that a 3-dimension tensor could degenerate existing metrics, and then illustrate how a generalized GoT characterizes a specific goal.

\textcolor{black}{\textbf{1) Degeneration to Existing Metrics:}}

\begin{figure}[htbp]
	\centering
	\includegraphics[angle=0,width=0.48\textwidth]{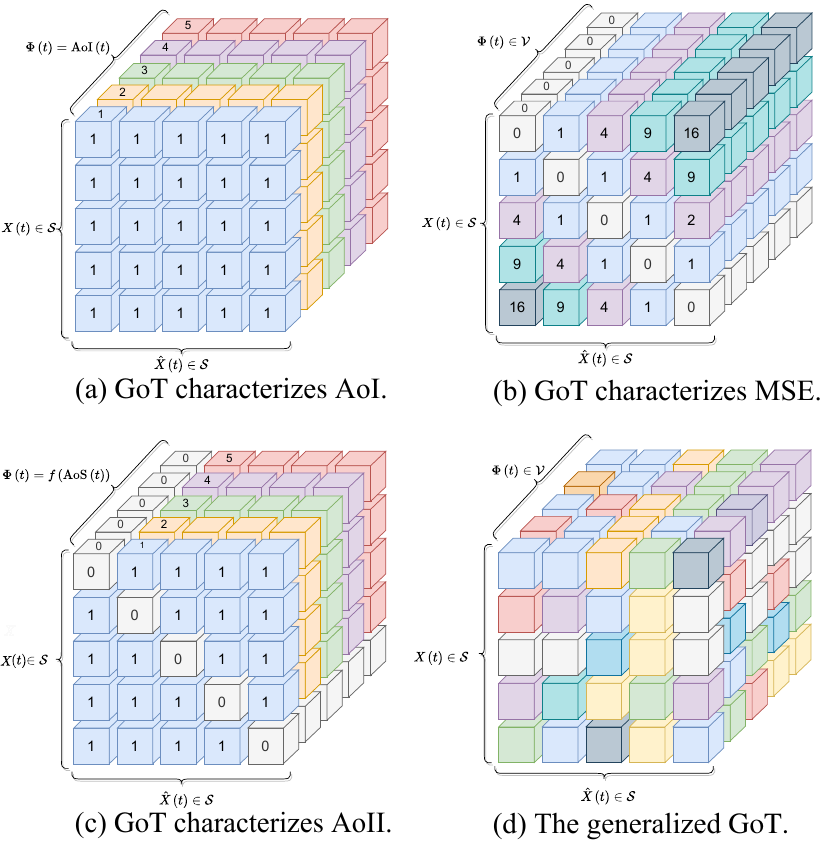}
	\caption{Examples of tensor-based metric visualizations.}\label{tensor}
\end{figure}

Fig. \ref{tensor}. (a) visualizes an example of the tensor to characterize AoI. Since AoI is \textbf{content-independent}, the values in the tensor only depend on the \textbf{content-independent} weighted coefficient $\Phi(t)$, with $\Phi(t)$ equals to AoI. Fig. \ref{tensor}. (b) visualizes a tensor to characterize MSE where $\mathcal{S}=\left\{0,1,2,3,4\right\}$. The MSE focuses on the squared error between $X\left(t\right)$ and $\hat{X}\left(t\right)$, while neglecting the \textbf{content-independent} weighted coefficient $\Phi(t)$. In this case, different $\Phi(t)$ produces the same tensor slice characterized by the squared error. Fig. \ref{tensor}. (c) visualizes an example for AoII. Since AoII is characterized in a multiplicative manner, a base slice exists in the AoII-based tensor. The base slice is obtained by the error gap function $g(X(t),\hat{X}(t))$, which is the first slice in front of us in Fig. \ref{tensor}. (c). A linear multiplication of the base slice could represent the other slices. The examples are not exhaustive. A similar approach could also be implemented to obtain a tensor to characterize VoI, AoS, UoI, and \emph{Cost of Actuation Error}.
\begin{figure*}[htbp]
	\centering
	\includegraphics[angle=0,width=1\textwidth]{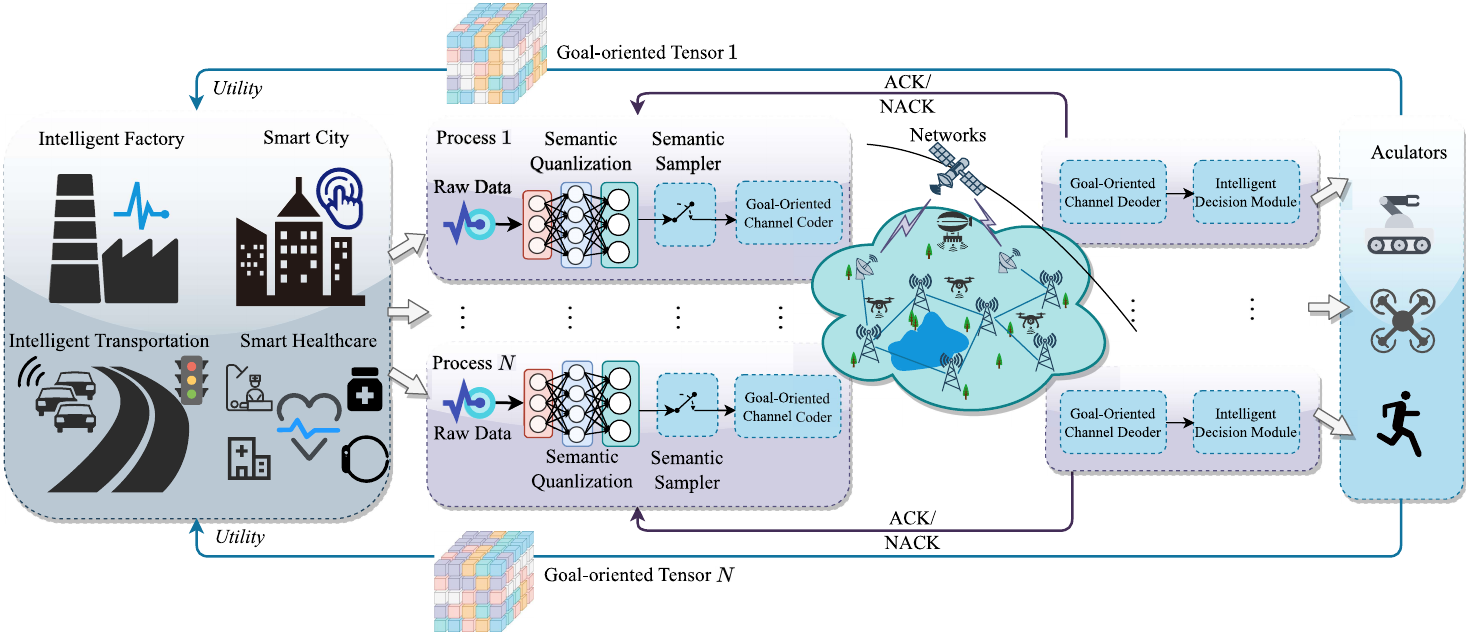}
	\caption{A sketch for the \textcolor{black}{semantics-empowered goal-oriented} networks.}\label{netowrk}
\end{figure*}

\textcolor{black}{\textbf{2) Generalized GoT}}

\textcolor{black}{A more generalized GoT is visualized in Fig. \ref{tensor}. (d). \textbf{First}, $X(t)$ is extended from content-agnostic raw status to semantics-aware status in the generalized GoT. This extension highlights the asymmetry of the mismatch cost between \(X\left(t\right)\) and \(\hat{X}\left(t\right)\), corroborating the principle of the \emph{Cost of Actuation Error}. The rationale of this extension could be illustrated through an autonomous driving situation: a false positive detection—perceiving a person where there is no one—may cause unnecessary braking but remains safe. In contrast, a false negative—the failure to detect an actual person—may lead to severe consequences. This highlights the need for an asymmetrical error function as an extension of the symmetrical one used in MSE and UoI. \textbf{Second,} the \textbf{content-independent} weighted coefficient $\Phi(t)$ is extended from multiplicative coefficient to context-aware representation. The role it plays in determining the tensor value is versatile, moving beyond a mere multiplicative method. Depending on the context, different end-to-end semantic discrepancies will result in varying costs, as depicted in Fig. \ref{tensor}. (d).}

\subsection{Constructing Generalized Goal-Oriented Tensor}\label{sectionC}
The values in the generalized GoT manifest the cost of E2E statuses mismatch under a specific context status. The values in the GoT should relate to the goal directly. Here a detailed method to tailor a GoT based on the goal is given. 

\noindent $\bullet$ \textbf{Step 1:} Clarifying the scenario and the goal. \textcolor{black}{In wireless accident monitoring and rescue systems, for example, the goal is to achieve effective monitoring and rescue such that the long-term average cost resulted by accident damages and rescue resources is minimized.}

\noindent $\bullet$ \textbf{Step 2:} Defining the \textcolor{black}{semantic status set} $\mathcal{S}$ and the \textcolor{black}{context status set} $\mathcal{V}$. \textcolor{black}{Both sets are constructed as finite collections of discrete semantic segments, with the semantic status $X(t)\in\mathcal{S}$ and $\Phi(t)\in\mathcal{V}$ determined through semantic quantization (refer to Section \ref{section4}).} For example, $\mathcal{S}$ could involve semantic fragments, whereas $\mathcal{V}$ could encapsulate different weather conditions surrounding the observed source.

\noindent $\bullet$ \textbf{Step 3:} Determining the decision strategy in terms of $\hat{X}\left(t\right)$. Status update will be converted to a decision update at the actuators. This decision will turn back to affect the goal-oriented effectiveness of the system.

\noindent $\bullet$ \textbf{Step 4:} Evaluating the specific cost. There are three types of costs: $i$) \textcolor{black}{the status inherent cost $\mathcal{C}_1(X(t),\Phi(t))$}, which represents the cost under $X\left(t\right)$ and $\Phi\left(t\right)$ in the absence of external influences; $ii$) \textcolor{black}{the decision gain $\mathcal{C}_2(\hat{X}(t))$}, which is based on the fact that a good $\hat{X}(t)$-dependent decision could reduce the severity; $iii$) \textcolor{black}{the decision cost $\mathcal{C}_3(\hat{X}(t))$}, which corresponds to the resource overhead due to a specific decision based on $\hat{X}(t)$

\noindent $\bullet$ \textbf{Step 5:} Calculating the GoT. The tensor value, given a specific triple-tuple $\langle X(t),\hat{X}(t),\Phi(t)\rangle$ and a determined decision mapping approach, is calculated by 
\begin{equation}
[\mathcal{C}_1(X(t),\Phi(t))-\mathcal{C}_3(\hat{X}(t))]^++\mathcal{C}_2(\hat{X}(t)),
\end{equation} where the ramp function $[\cdot]^+$ ensures that any decision gain $\mathcal{C}_3$ will not reduce the cost below $0$.

\subsection{\textcolor{black}{How does GoT Address the Semantics and Effectiveness?}}
\textcolor{black}{\textbf{\textit{1) Semantics:}} The GoT addresses semantics from dual perspectives. First, it extends the distance of E2E meaningless raw data mismatch to the cost of context-specific E2E semantic mismatches. The interpretation of semantics as meaning is thus addressed. Second, the E2E semantic mismatch cost serves as a gauge of the \emph{relevance} of information (or \emph{relevance} of semantics as $X(t)$ represents semantic status). A higher context-specific mismatch cost indicates greater semantic relevance and vice versa, emphasizing the role of semantics in determining \emph{relevance}.}

\textcolor{black}{\textbf{\textit{2) Effectiveness:}} The tensor value within GoT is determined by both the E2E semantic mismatch and the decision strategy, representing the real-world cost due to these mismatches. Through GoT optimization, one can enhance decision-making and diminish significant E2E semantic inconsistencies. As a result, the long-term real-world cost is effectively reduced, highlighting GoT's focus on effectiveness.}

\section{A Holistic Framework for {Goal-Oriented Semantic} Communication}\label{section4}

This section introduces a comprehensive goal-oriented semantic framework, depicted in Fig. \ref{netowrk}, with the GoT acting as the central metric. The modules of our framework are elaborated as follows:

\textbf{Semantic Quantization.}
Status update packets are generated when the sensors collect continuous raw data. The types of metadata are various, e.g., a piece of video/speech, a process of temperature variations, a fragment of moving track, or even their multi-modal combinations. \textcolor{black}{Raw data are fed into a semantic quantizer for semantic extractions, producing semantics $X(t)$ and context $\Phi(t)$. This quantizer converts continuous, non-meaningful metadata into discrete semantic pieces}. Leveraging DL-based methods, this transformation reduces the data volume per packet and facilitates the subsequent sparse semantics-aware sampling design.

\textbf{Sparse Semantics-Aware Sampling.}
{Context status $\Phi(t)$, semantic status $X(t)$, and the feedback signal, \emph{i.e.}, ACK or NACK are fed into the sampler to decide whether the current status should be sampled and transmitted. In our framework, $\Phi(t)$ can illustrate the context surrounding the perceived source and is a sensor-derived meaningful perception.} Only updates that facilitate goal achieving are transmitted, while others are filtered. In an ideal design, sparse semantics-aware sampling that fulfills the goal can be achieved.

\textbf{Goal-oriented Coding.}
{Traditionally, channel coding strategies aim for perfect reconstructions. This method overlooks the delay-caused E2E status mismatch and its cost to the goal-oriented actuation. To address this issue, E2E decoding errors that may result in high costs can be assigned larger code distances. This allocation strategy reduces the probability of costly mismatches occurring. Conversely, decoding errors associated with lower costs can be allotted shorter code distances. By adopting this module, the system effectiveness can be further enhanced.}

\textbf{Intelligent Decision Making.}
This process indicates the end of the \emph{life cycle} of information, which intelligently transforms an estimated status update into a particular utilizable decision update. The decision could be executed by actuators, such as an automobile, a remote-controlled UAV, a robotic arm, or well-trained people. The actuators transform the decisions into the ultimate \textit{\textbf{effectiveness}}, which generally turns back to exert positive influences on the evolutionary process of the source.

\section{Case Study: Goal-oriented Sparse Semantic Sampling}\label{sectionV}
\begin{figure}[htbp]
	\centering
	\includegraphics[angle=0,width=0.48\textwidth]{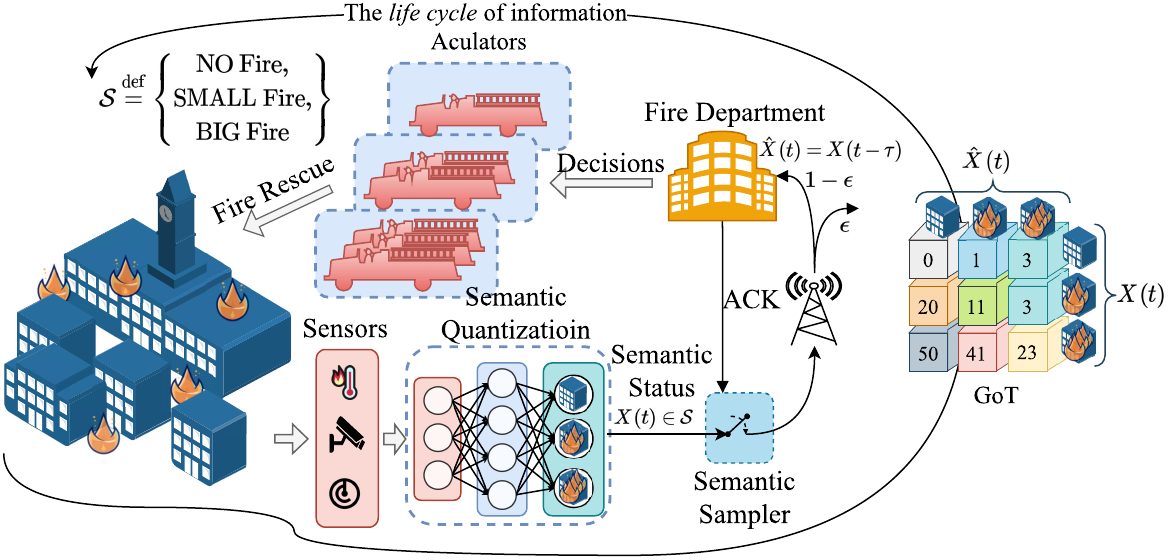}
	\caption{Case study: A real-time wireless fire monitor and rescue system. }\label{casestudy}
\end{figure}
This section considers the real-time wireless fire monitor and rescue system in Fig. \ref{casestudy} as a case study. The goal in Fig. \ref{casestudy} is to minimizing the long-term average economic losses and resource consumption associated with fire and rescue induced by fire and rescue. As an initial exploration, our focus in this article is on the design of sparse semantics-aware sampler. The sampler should decide which semantics to sample and transmit, such that the rescue is timely and effective to minimize the long-term real-world costs induced by fire and rescue. This E2E status mismatch cost could be recorded by GoT, and thus to achieve the goal is equivalent to minimizing the long-term average GoT. 

\begin{figure}[htbp]
	\centering
	\includegraphics[angle=0,width=0.42\textwidth]{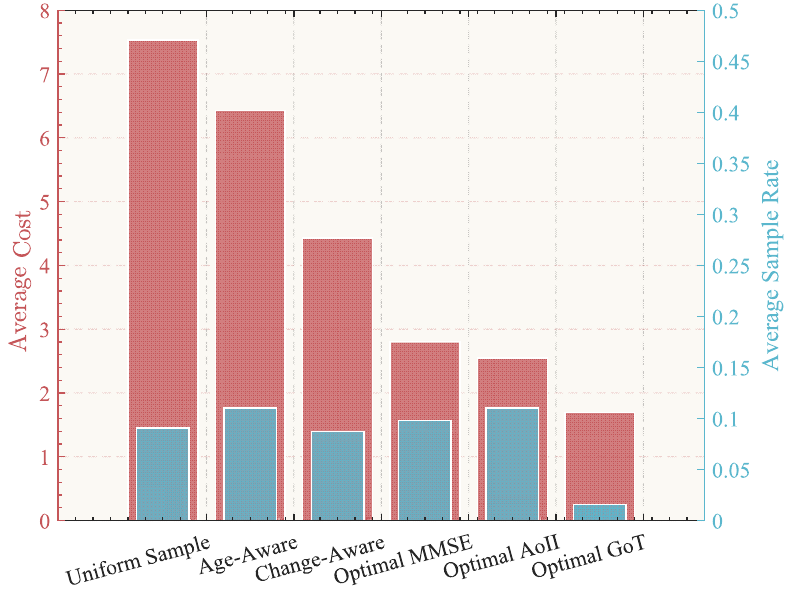}
	\caption{The performance comparisons among different sample policies. Here the goal is to reduce the \textcolor{black}{long-term average cost} caused by fire and rescue. }\label{simulation}
\end{figure}

As a case study, we evaluate a basic system where the context status $\Phi(t)$ is constant. The semantic status captures three fire severity levels. In this way, the $3\times3$ GoT is visualized at the right-hand side of Fig. \ref{casestudy}, signifying that the semantics ``NO FIRE'' estimated as ``NO FIRE'' will consume zero real-world cost, while ``SMALL FIRE'' estimated as ``NO FIRE'' will incur the 20 real-world cost per time slot, etc. We assume perfect semantic quantization. The semantic sampler decides when and what to \textcolor{black}{sample and transmit} a status through an $\epsilon$-erasure channel to the fire department. Firefighters (actuators) are dispatched based on the latest fire levels. The time is slotted and the source is modeled as a controlled Markov Source. The transition probabilities of the fire levels depend on the firefighters assigned. In such a system, we compare the following six sampling policies in terms of goal-oriented effectiveness and sampling rate:

\noindent $\bullet$ \textbf{Uniform.} Sampling is triggered periodically, which is independent of the packet's content.

\noindent $\bullet$ \textbf{Age-aware.} Sampling is conducted once the AoI reaches a pre-defined threshold, which is content-agnostic.

\noindent $\bullet$ \textbf{Change-aware.} Sampling is triggered when the source changes, which is content-aware.

\noindent $\bullet$ \textbf{Optimal MMSE.} Sampling is designed to minimize long-term average MSE, which is content-aware.

\noindent $\bullet$ \textbf{Optimal AoII (E2E-Semantics).} Sampling is designed to minimize long-term average AoII, which holds consistent with the \emph{E2E-semantics policy} proposed in \cite{Nikolaos}. Specifically, sampling is triggered once an E2E status mismatch arises.

\noindent $\bullet$ \textbf{Optimal GoT.} Sampling is designed to minimize \textcolor{black}{long-term average GoT, which is equivalent to minimizing the long-term average cost resulted by personnel casualties, equipment and building damages, as well as manpower or resources consumption in the real world. The sampling policy is solved resorting to the Markov Decision Process (MDP)}.

Fig. \ref{simulation} showcases the simulated results that compare the performance of the discussed sampling policies. The performance is evaluated by two indicators: long-term average cost, reflecting the cumulative real-world implications of fire and rescue, and average sample rate, indicating communication overhead. The primary goal is to reduce the long-term average cost, and as the Fig. \ref{simulation} reveals, the GoT-optimal policy excels in this, marking its inherent alignment with our goal. This policy also demonstrates an efficient balance between reduced communication overhead (sample rate) and effectiveness, suggesting a semantics-aware, goal-driven sampling method. For instance, the semantic status ``BIG FIRE'' is always more important due to the fact that it will incur more severe cost. These sampling policies are all event-triggered; therefore, the average cost and sampling rate are derived concurrently. From Fig. \ref{simulation}, the proposed GoT-optimal policy stands out by emphasizing the most goal-relevant, semantics-rich data while ensuring minimum costs.

In a general case, the most important thing is to tailor the GoT to align the specific scenarios and goals. This process motivates accessing the real-world costs of context-specific status mismatch categories. With the tailored GoT in hand, the goal-relevant sparse sampling could be developed similarly.

\section{Conclusions and Future Challenges}\label{sectionVI}
The \emph{relevance} of information plays a critical role in interpreting `\emph{semantics}'. Following this avenue, this work moves further by proposing a new goal-oriented metric and envisioning a holistic goal-oriented architecture. The proposed GoT provides a unified and extensible solution to characterize the ultimate goal of communications, which provides a solution to the {{effectiveness}} problem. The proposed goal-oriented network architecture demonstrates its great superiority in substantially alleviating the communication burden of the next-generation Internet of Everything (IoE) networks. A preliminary instantiated case study is demonstrated to address the challenge of sparse semantic sampling presented in \cite{uysal2022semantic} and \cite{Nikolaos}. 

Towards the promising avenue of research, some interesting open challenges have been discussed in \cite{uysal2022semantic} and\cite{Nikolaos}, which also aligns with our proposed framework. Here we complement some open challenges that have not yet been presented: 

\textbf{Communication Networks with Heterogeneous Goals.} The intelligent factory, smart cities, intelligent transportation, and smart healthcare are typical real-time applications for the future IoE networks. Sensors in this network collect terabytes of metadata every second, placing a great deal of pressure on the constrained communication resources. As such, it is imperative to slim down the data from its generation. An initial architecture to address this issue is shown in Fig. \ref{netowrk}, where the E2E heterogeneous goals are characterized by different GoTs with diversified sizes and values, and the semantic sampling and coding are optimized from a system perspective.

\textbf{Goal-oriented physical (PHY) layer techniques.}
To achieve a specific goal under constrained resources, the bit-by-bit reconstruction is so energy-intensive and low-efficiency that it could not perceive the \emph{priority} of information in terms of the ultimate effectiveness to adaptively allocate the limited resources. In this regard, new goal-oriented paradigms at the PHY could be explored to further improve communication efficiency. In particular, the goal-oriented channel coding and decoding algorithms, the goal-oriented retransmission and feedback mechanism, the multi-user power allocation mechanism, the goal-oriented modulation and signal shaping are some promising candidates. By this means, the grand vision beyond the traditional paradigm is that the future communication is not designed to only reduce the error probability in an average manner but to circumvent the severe errors.

\textbf{Goal-oriented Perception, Communication, Computation, and Control Co-design }.
The \emph{life cycle} of information is closely affiliated with the processes of perception, communication, computation, caching, and control. Therefore, a true leap forward can be achieved by merging these modules together for the co-design optimizations. Particularly, the multi-modal sensors consecutively perceive high-resolution raw data, the computation leads to precise semantic extractions and effective intelligent decisions, the communication serves as a \emph{priority}-aware information pipeline for information disseminating, and the actuators executes the control commands precisely, so as to transform the information into practical usage. 

\bibliographystyle{IEEEtran}
\bibliography{reference}
%
\vspace{-1.5cm}
\IEEEbiographynophoto{Aimin Li} received his B.E. from Harbin Institute of Technology Shenzhen (HITSZ) in 2020, where he was awarded the highest honor of Undergraduate Thesis. He is currently a Ph.D student at HITSZ. He has served as a Reviewer for the IEEE Trans. Wireless Communications and the IEEE Trans. Neural Networks and Learning Systems. His current research interests include age of information and goal-oriented semantic communications.
\vspace{-1.5cm}
\IEEEbiographynophoto{Shaohua Wu} received the Ph.D.
degree in communication engineering from Harbin
Institute of Technology, Harbin, China, in 2009. is a Full Professor at HITSZ and Peng Cheng Laboratory. He was also a Visiting Researcher at the University of Waterloo's BBCR in 2014-2015. His current research interests include wireless image/video transmission, space communications,
advanced channel coding techniques, and B5G wireless transmission technologies. He has authored or co-authored over 50 Chinese patents and 200 articles, two of which have received the best paper awards. 
\vspace{-1.5cm}
\IEEEbiographynophoto{Siqi Meng} received his B.E. from Harbin Institute of Technology Shenzhen (HITSZ) in 2021. He is currently a Ph.D. student in HITSZ. His research interests include age of information and semantic communications.
\IEEEbiographynophoto{Rongxing Lu} received the Ph.D. degree from the Department of Electrical and
Computer Engineering, University of Waterloo, Waterloo, ON, Canada, in 2012. He is a Mastercard IoT Research Chair, a University Research Scholar, an Associate Professor with the Faculty of Computer Science (FCS), University of New Brunswick (UNB), Fredericton, NB, Canada. Before that, he worked as an Assistant Professor with the School of Electrical and Electronic Engineering, Nanyang Technological University, Singapore, from 2013 to 2016. He worked as a Postdoctoral Fellow with the University of Waterloo from 2012 to 2013. His research interests include applied cryptography, privacy enhancing technologies, and IoT-Big Data security and privacy. Dr. Lu won the 8th IEEE Communications
Society (ComSoc) Asia–Pacific Outstanding Young Researcher Award in
2013. He currently serves as the Chair for the IEEE ComSoc Communications and Information Security Technical Committee, and the Founding Co-Chair for the IEEE TEMS Blockchain and
Distributed Ledgers Technologies Technical Committee.
\vspace{-1cm}
\IEEEbiographynophoto{Sumei Sun} is a Principal Scientist, Distinguished Institute Fellow, and Acting Executive Director of the Institute for Infocomm Research (I2R), Agency for Science, Technology, and Research (A*STAR), Singapore. She is also holding a joint 
appointment with the Singapore Institute of Technology, and an adjunct appointment 
with the National University of Singapore, both as a full professor. Her current 
research interests are in next-generation wireless communications, cognitive 
communications and networks, industrial internet of things, communications-computing-control integrative design, joint radar-communication systems, and signal 
intelligence. She is Editor-in-Chief of the IEEE Open J. Vehicular Technology, and 
Steering Committee Chair of the IEEE Trans. Machine Learning in 
Communications and Networking. She is also Member-at-Large of 
the IEEE Communications Society, and Board of Governors member of the IEEE Vehicular Technology Society.
\vspace{-1cm}
\IEEEbiographynophoto{Qinyu Zhang} received his
B.E. from Harbin Institute of Technology (HIT), Harbin, China, in 1994, and the Ph.D. degree in biomedical and electrical engineering from the University of Tokushima, Tokushima, Japan, in 2003. He was an Assistant Professor with the University
of Tokushima from 1999 to 2003. He has been with
HITSZ since 2003, where
he is currently a Full Professor and serves as the Vice
President. His research interests include aerospace
communications and networks, and wireless communications and networks. He is the Founding Chair of the \emph{IEEE Communications Society Shenzhen Chapter}. He has been awarded the National Science Fund for Distinguished
Young Scholars, Young and Middle-Aged Leading Scientist of China and
the Chinese New Century Excellent Talents in University, and obtained three
scientific and technological awards from governments.

\end{document}